\documentclass[aps,prb,reprint,groupedaddress]{revtex4-2}

\usepackage{graphicx}
\usepackage{textcomp}
\usepackage{xcolor}
\usepackage[colorlinks=true,linkcolor=blue,anchorcolor=blue,citecolor=blue,urlcolor=blue]{hyperref}

\begin{document}
\title{Probing Sign-Changing Order Parameters via Impurity States in unconventional superconductors: Implications for La$_3$Ni$_2$O$_7$ Superconductors with interlayer pairing}
\author{Junkang Huang$^{1,2}$}
\author{Z. D. Wang$^{3,4}$}
\email{zwang@hku.hk}
\author{Tao Zhou$^{1,2}$}
\email{tzhou@scnu.edu.cn}

\affiliation{$^1$Guangdong Basic Research Center of Excellence for Structure and Fundamental Interactions of Matter, Guangdong Provincial Key Laboratory of Quantum Engineering and Quantum Materials, School of Physics, South China Normal University, Guangzhou 510006, China\\
	$^2$Guangdong-Hong Kong Joint Laboratory of Quantum Matter, Frontier Research Institute for Physics, South China Normal University, Guangzhou 510006, China\\
	$^3$HK Institute of Quantum Science $\&$ Technology and Department of Physics, The University of Hong Kong, Pokfulam Road, Hong Kong, China\\
	$^4$Hong Kong Branch for Quantum Science Center of Guangdong-Hong Kong-Macau Great Bay Area, 3 Binlang Road, Shenzhen, China
}

\begin{abstract}
	
Motivated by the desire to investigate the fundamental relationship between impurity-induced states and the sign change of the superconducting order parameter, as well as to explore the impurity effects in Ruddlesden-Popper nickelate superconductors with interlayer pairing, we employ the $T$-matrix approach to study single impurity scattering in unconventional superconductors. Our work focuses on two distinct pairing scenarios: intralayer $d$-wave pairing and interlayer $s$-wave pairing. For systems with intralayer $d$-wave pairing, we establish an intrinsic connection between the $d$-wave pairing symmetry and the emergence of mid-gap resonant states. Through a combination of analytical derivations and numerical simulations, we demonstrate that the appearance of in-gap states is directly linked to the sign reversal of the order parameter along the Fermi surface. In interlayer pairing systems, our results reveal the presence of pronounced resonant peaks, which can also be attributed to the sign-changing nature of the order parameter. We further extend our analysis to the bilayer nickelate superconductor La$_3$Ni$_2$O$_7$, providing a theoretical investigative of impurity effects in this material. Our findings not only elucidate the complex interplay between pairing symmetries and impurity-induced states in unconventional superconductors but also offer a powerful tool for probing the pairing mechanisms in nickelate-based high-temperature superconductors. This work lays the groundwork for future experimental and theoretical investigations into the unique electronic properties of these emerging materials.


\end{abstract}
\maketitle

\section{Introduction}

The single impurity effect has been extensively studied in unconventional superconducting systems, including cuprate high-temperature superconductors \cite{Pan2000,RevModPhys.78.373,PhysRevB.67.094508,PhysRevLett.86.296,PhysRevLett.96.097004,PhysRevLett.89.287002}, iron-based superconductors \cite{PhysRevB.83.214502,PhysRevB.96.220507, PhysRevB.80.064513,PhysRevB.80.104504,PhysRevLett.103.186402,PhysRevB.94.144512, PhysRevB.81.014524,PhysRevB.88.220509,PhysRevB.99.014509,JPSJ.78.084718,PhysRevB.79.054529,PhysRevB.84.134507}, as well as other superconducting systems \cite{PhysRevB.78.035414, s11467-021-1056-y, PhysRevB.109.054504, PhysRevB.108.144508, PhysRevB.108.174501, PhysRevB.98.024515, PhysRevLett.110.020401, PhysRevA.87.013622, PhysRevB.88.205402, PhysRevB.89.214506, PhysRevB.95.140202, PhysRevLett.105.046803,Zha_2017,Guo2017,PhysRevB.91.165142,PhysRevB.101.064510,PhysRevB.104.235426,PhysRevB.68.224502,PhysRevB.88.094514}. These studies have been facilitated by theoretical explorations and experimental observations through local density of states (LDOS) spectra and scanning tunneling microscopy (STM) measurements.

In the case of cuprate high-temperature superconductors with $d$-wave pairing symmetry, a noteworthy discovery is the emergence of a strong resonant state induced by a point impurity at the middle point of the energy gap, commonly known as the mid-gap state \cite{Pan2000,RevModPhys.78.373}. The presence of the mid-gap state is attributed to the process of consecutive Andreev reflections \cite{PhysRevLett.72.1526}. For iron-based superconductors, it has been proposed that impurity effects can serve as a probe for the pairing symmetry, with impurity-induced in-gap states potentially confirming the sign-reversing nature of the superconducting gap across different Fermi surface pockets \cite{PhysRevB.83.214502,PhysRevB.96.220507,PhysRevB.80.064513, PhysRevB.94.144512, PhysRevLett.103.186402,PhysRevB.81.014524,PhysRevB.88.220509,PhysRevB.99.014509,JPSJ.78.084718,PhysRevB.79.054529}.

Theoretically, the impurity effects in unconventional superconducting systems can be analyzed using the $T$-matrix method, which suggests that in-gap states are associated with singularities in the $T$-matrix denominator \cite{RevModPhys.78.373}. However, the precise relationship between the mid-gap state and $d$-wave pairing symmetry within the $T$-matrix framework remains to be in-depth explored. In particular, the link between in-gap states and the sign change of the order parameter is not clearly established, indicating a need for further research to clarify these interactions and their impact on system behavior.

Recently, the bilayer nickel-based superconducting material La$_3$Ni$_2$O$_7$ has emerged as a novel high-temperature superconductor, garnering significant research interest \cite{s41586-023-06408-7,PhysRevLett.131.126001,Ko2024,zhou2024ambient}. The pairing symmetry in this compound remains an open question, with both $s_{\pm}$ and $d$-wave pairing symmetries being proposed \cite{Wang_2024,2024arXiv241211429G}. Additionally, several researchers have proposed that nickel-based high-temperature superconducting materials may exhibit dominant interlayer pairing characteristics \cite{rs.3.rs-3901266/v1, j.scib.2024.07.030, PhysRevB.108.L201108, PhysRevB.110.094509, PhysRevB.108.174511, PhysRevB.110.L041111, PhysRevLett.132.146002, 0256-307X/41/5/057403,PhysRevB.111.094504,huang2025spmpairing}. Therefore, expanding the investigation of single impurity scattering effects to systems dominated by interlayer pairing could be instrumental in elucidating the pairing symmetry of nickel-based superconductors.

In this paper, we revisit the effects of single impurity scattering in high-temperature superconducting materials using the $T$-matrix method. We consider two scenarios: intralayer pairing and interlayer pairing. For high-temperature superconductors with intralayer pairing, we elucidate why cuprates with $d$-wave pairing invariably lead to impurity-induced mid-gap states. We establish an intrinsic connection between the sign change of the order parameter and the emergence of in-gap states through the $T$-matrix approach. For $s$-wave superconductors with interlayer pairing, our results indicate that impurities induce strong in-gap resonant states, which can also be attributed to the sign reversal of the order parameter.

The structure of our paper is outlined as follows: In Section II, we introduce the Hamiltonian and the methodologies employed in our study. In Sec.III, we aim to establish a fundamental link between the impurity induced bound states and the sign changing of the order parameter along the normal state Fermi surface. 
In Sec.IV, we study the impurity effect of the bilayer nickelate superconductor with the interlayer pairing. We conclude with a brief summary of our findings in Section V.

\section{\label{sec:Model}Model and Formalism}
Our investigation commences with a multilayer tight-binding model that incorporates superconducting pairing terms, formulated as follows:

\begin{eqnarray}
	H= \sum_{{\bf k}l\sigma}\varepsilon_{\bf k} c_{{\bf k}l\sigma}^{\dagger}c_{{\bf k}l\sigma} - \sum_{\bf k}t_{\perp} \left(c_{{\bf k}1\sigma}^{\dagger}c_{{\bf k}2\sigma} + H.c.\right) + H_{\Delta},
\end{eqnarray}
where \(\varepsilon_{\mathbf{k}} = -2t\left( \cos k_x + \cos k_y \right) - \mu\) with \(t\) being the nearest neighbor hopping constant. Here, \(l\) and $\sigma$ denote the layer and spin indices, respectively. \(\mu\) is the chemical potential, and \(t_{\perp}\) describes the interlayer hopping constant for the bilayer system.

The superconducting paring term $H_{\Delta}$ is given by
\begin{eqnarray}
	H_{\Delta} = \sum_{{\bf k}ll^\prime}(\Delta^{l,l^\prime}_{\bf k} c_{{\bf k}l\uparrow}^{\dagger}c_{{-\bf k}l^\prime\downarrow}^{\dagger} + H.c.),
\end{eqnarray}
where $l=l^\prime$ and $l\neq l^\prime$ represent the intralayer pairing and interlayer pairing, respectively. 

The Hamiltonian Eq. (1) can be expressed in matrix form as $H=\sum_{\bf k}\hat{\Psi}_{\bf k}^\dagger \hat{M}_{\bf k} \hat{\Psi}_{\bf k}$, where $\hat{M}_{\bf k}$ is a $2N_l \times 2N_l$ matrix ($N_l$ is the number of layers). 
For a bilayer system ($N_l=2$), the column vector $\hat{\Psi}_{\bf k}$ is defined as $\hat{\Psi}_{\bf k}=(c_{1{\bf k}\uparrow},c_{2{\bf k}\uparrow}, 
c^\dagger_{1{\bf -k}\downarrow},
c^\dagger_{2{\bf -k}\downarrow})^T$, and the matrix $\hat{M}_{\bf k}$ takes the form,
\begin{eqnarray}
\hat{M}_{\bf k} = \left( \begin{array}{cccc}
{\varepsilon_{\mathbf{k}}}&{-t_{\perp} }&\Delta^{11}_{\bf k} & \Delta^{12}_{\bf k}\\
-t_{\perp}&{\varepsilon_{\mathbf{k}}}&\Delta^{21}_{\bf k}&\Delta^{22}_{\bf k}\\
\Delta^{11*}_{\bf k}&\Delta^{21*}_{\bf k}&-\varepsilon_{\mathbf{k}}&t_{\perp}\\
\Delta^{12*}_{\bf k}&\Delta^{22*}_{\bf k}&t_{\perp}&-\varepsilon_{\mathbf{k}}\
\end{array} \right).
\end{eqnarray}
Diagonalizing $\hat{M}_{\bf k}$ via $\hat{D}_{\bf k}=
\hat{R}_{\bf k}^{\dagger} \hat{M}_{\bf k} \hat{R}_{\bf k}$ (where $\hat{R}_{\bf_k}$ is a unitary matrix and $\hat{D}_{\bf k}$ is diagonal with eigenvalues $E_n({\bf k})$),  
the bare Green's function matrix for a clean system can then be obtained, with the elements being defined as
\begin{eqnarray}
	G_{0ij}\left({\bf k},\omega\right) =\sum_n \frac{u_{in}({\bf k})u^*_{jn}({\bf k})}{\omega - E_n\left({\bf k}\right) + i\Gamma},
\end{eqnarray}
where $\Gamma$ is a small constant and $u_{in}({\bf k})$ is the elements of the matrix $\hat{R}_{\bf k}$.

Considering a single impurity at the site $\left(0,\space 0\right)$ of layers 1, the $T$-matrix can be expressed as \cite{RevModPhys.78.373}
\begin{eqnarray}
	\hat{T}\left(\omega\right) = \frac{\hat{U}}{\hat{I} - \hat{U}\hat{G}_{0}\left({\bf r},{\bf r},\omega\right)},
\end{eqnarray}
where \(\hat{I}\) is the identity matrix, \(\hat{U}\) is a diagonal matrix with non-zero elements \(U_{11} = V_{\text{imp}}\) and \(U_{1+N_l,1+N_l} = -V_{\text{imp}}\), and \(\hat{G}_{0}\left(\mathbf{r},\mathbf{r'},\omega\right)\) is the Fourier transform of \(\hat{G}_0\left(\mathbf{k},\omega\right)\) with
$
\hat{G}_{0}\left(\mathbf{r},\mathbf{r'},\omega\right) = \frac{1}{N} \sum_{\mathbf{k}} \hat{G}_0\left(\mathbf{k},\omega\right) e^{i\mathbf{k} \cdot (\mathbf{r}-\mathbf{r'})}$.

The full Green's function in the presence of a single impurity is calculated by:
\begin{equation}
\hat{G}\left(\mathbf{r},\mathbf{r'},\omega\right) = \hat{G}_0\left(\mathbf{r},\mathbf{r'},\omega\right) + \hat{G}_0\left(\mathbf{r},0,\omega\right) \hat{T}\left(\omega\right) \hat{G}_0\left(0,\mathbf{r'},\omega\right).
\end{equation}

The LDOS at layer \(l\) and site \(\mathbf{r}\) can be calculated using the full Green's function:
\begin{equation}
\rho_{l}\left(\mathbf{r},\omega\right) = -\frac{1}{\pi} \text{Im} \left[ G_{ll}\left(\mathbf{r},\mathbf{r},\omega\right) + G_{l+N_l,l+N_l}\left(\mathbf{r},\mathbf{r},-\omega\right) \right].
\end{equation}

In the following presented results, we set the nearest-neighbor hopping parameter \(t\) as the energy unit with \(t = 1\). Other parameters are set as \(\Gamma = 0.01\) and \(V_{\text{imp}} = 20\).

\section{Relation between Impurity states and the sign change of the order parameter}

The relationship between non-magnetic impurity-induced bound states and the sign-changing nature of the superconducting order parameter along the Fermi surface represents a critical area of research in unconventional superconductivity. Pioneering theoretical studies, grounded in the principles of Andreev reflection, predicted the formation of mid-gap states at the surfaces of 
$d$-wave superconductors \cite{PhysRevLett.72.1526}. These states were directly tied to the sign reversal of the $d$-wave order parameter across different momentum directions, serving as a hallmark feature of $d$-wave pairing symmetry. Subsequent work further elucidated that the emergence of impurity-induced mid-gap states in  
$d$-wave superconductors could similarly be attributed to Andreev reflection processes, reinforcing the connection between sign-changing order parameters and localized in-gap states~\cite{PhysRevLett.96.097004}. While Andreev reflection provides a qualitative physical picture for this correlation, the reliance on such arguments remains largely intuitive, necessitating a rigorous numerical and analytical validation-particularly through the $T$-matrix formalism-to quantitatively link impurity-induced states to the symmetry-imposed sign changes in the order parameter.

In iron-based superconductors, theoretical studies have proposed impurity-induced bound states as spectroscopic tools to detect gap sign variations, with the mechanism again tied to Andreev reflection-mediated interference effects~\cite{PhysRevB.80.064513,PhysRevLett.103.186402,PhysRevB.80.104504}. Analogous phenomena are observed in some other systems with unconventional superconducting pairing, where gap sign changes correlate with impurity states~\cite{s11467-021-1056-y, PhysRevB.109.054504}. However, the predominance of qualitative models underscores the urgency of a systematic $T$-matrix framework to resolve ambiguities and establish universality. Such an approach would not only clarify whether the impurity-signature connection reflects a fundamental principle but also deepen insights into how impurity potentials interact with unconventional pairing states. By combining numerical simulations with analytical $T$-matrix solutions, researchers could disentangle the roles of gap symmetry in bound-state formation. This would provide a robust foundation for probing pairing symmetries in emerging unconventional superconductors.

\begin{figure}
	\centering
	\includegraphics[width = 8.5cm]{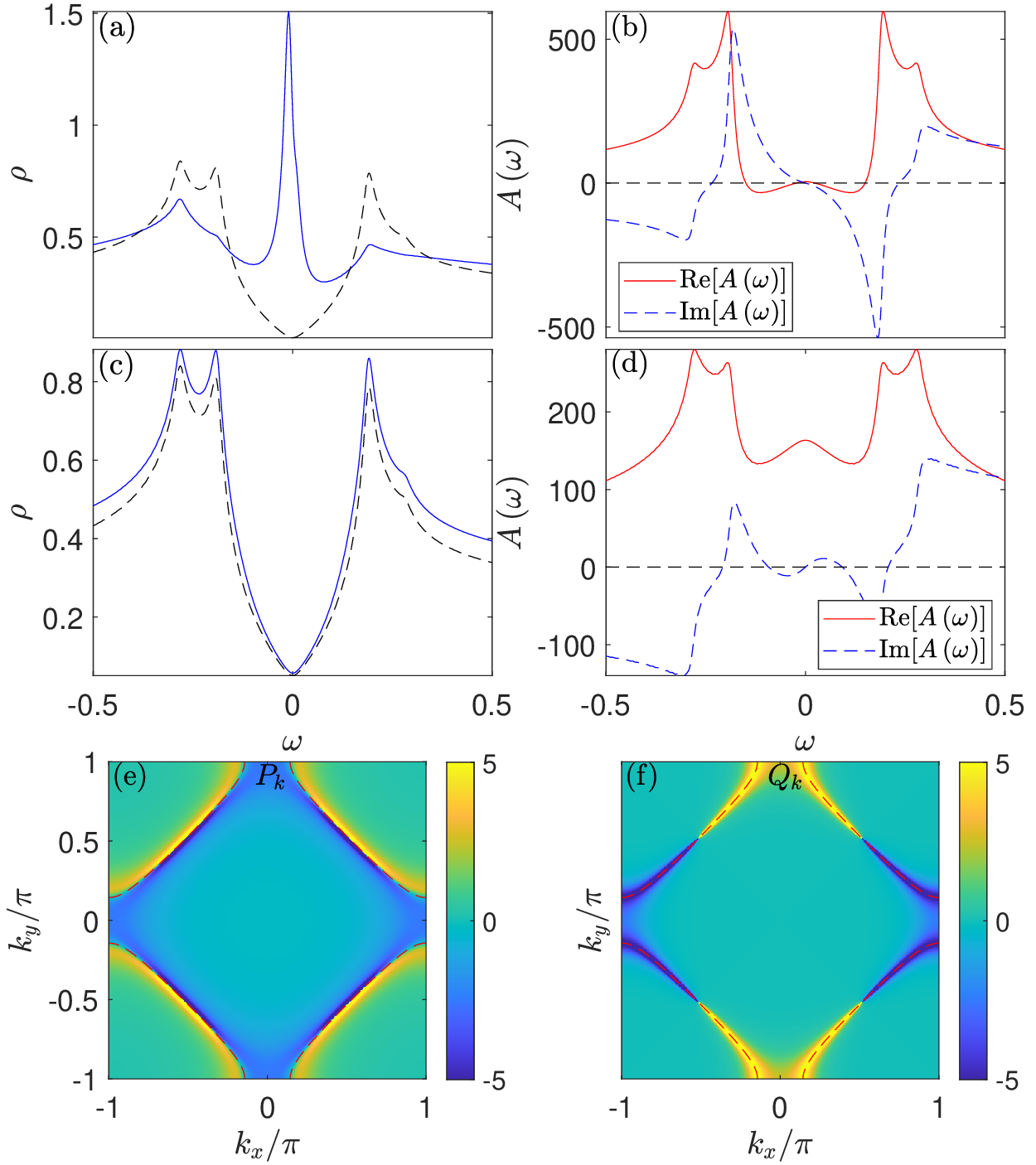}
	\caption{\label{fig1} (a) Solid line: LDOS spectrum at the nearest neighbor site of the impurity site for a $d_{x^2-y^2}$-wave pairing scenario. The dashed line represents the bare LDOS spectrum in the absence of the impurity. (b) Real and imaginary components of the function $A(\omega)$ for the $d$-wave pairing symmetry. Panels (c) and (d) correspond to panels (a) and (b), respectively, but for the sign-unchanged pairing with $\Delta_{\mathbf{k}} = |\Delta_0 (\cos k_x-\cos k_y)/2|$. Panels (e) and (f) display the intensity distribution maps for the functions $P_{\mathbf{k}}$ and $Q_{\mathbf{k}}$, respectively. The dashed lines in Panels (e) and (f) indicate the normal state Fermi surface.}
\end{figure}

In this section, motivated by above consideration, we study the impurity effect with a minimum single-layer model by setting \( l \equiv 1 \) and \( t_{\perp} = 0 \) in Eq.~(1). Initially, we re-exhibit the numerical results for the impurity effect on $d_{x^2-y^2}$-wave superconductors, taking \( \Delta_{\mathbf{k}} = \Delta_0 (\cos k_x-\cos k_y)/2 \) with \( \Delta_0 = 0.2 \). The LDOS spectra without the impurity and near an impurity are plotted in Fig.~\ref{fig1}(a). As observed, without the impurity, a $V$-shaped spectrum is obtained due to the existence of nodal points of the $d$-wave pairing symmetry.
A sharp mid-gap resonant peak emerges near the Fermi energy in the presence of the impurity. This mid-gap resonant peak can be explained through the denominator of the $T$-matrix, $A(\omega)$, with 
\[ A(\omega) = \mathrm{Det}\left[ \hat{I} - \hat{U}\hat{G}_{0}\left(\mathbf{r},\mathbf{r},\omega\right) \right]. \]
The impurity-induced resonant peaks emerge when both the real part and the imaginary part of $A(\omega)$ are zero at a certain low energy. For the $d$-wave pairing symmetry, the real and imaginary parts of $A(\omega)$ as a function of the energy $\omega$ are shown in Fig.~\ref{fig1}(b). As seen, the imaginary part of $A(\omega)$ crosses the zero axis at $\omega=0$, and the real part also crosses the zero axis near the Fermi level, leading to the mid-gap resonant peak.

It is widely believed that the mid-gap state presented in Fig.~\ref{fig1}(a) arises from the sign change of the $d$-wave order parameter. To verify this conclusion numerically, we consider the sign-unchanged order parameter with \( \Delta_{\mathbf{k}} = |\Delta_0 (\cos k_x-\cos k_y)/2| \). 
It is important to clarify that this specific configuration of the order parameter is not representative of any actual material; rather, it serves as a theoretical construct designed to highlight how the sign change in order parameters influences the impurity effect.
The LDOS spectra, both in the absence and presence of an impurity with this order parameter form, are illustrated in Fig. \ref{fig1}(c). The corresponding real and imaginary parts of $A(\omega)$ are displayed in Fig.~\ref{fig1}(d). As is seen, although the base LDOS spectrum without the impurity is the same as that of the $d$-wave superconductor, the spectrum in the presence of the impurity is significantly different. In this case, no in-gap structure exists. This absence of in-gap structure aligns with the numerical calculations of the denominator of the $T$-matrix, $A(\omega)$, where the real part of $A(\omega)$ is significantly far from the zero axis, as seen in Fig.~\ref{fig1}(d). These numerical results indicate that the impurity effect is indeed sensitive to the phase of the order parameter and thus can be used to detect the sign change of the order parameter.

To delve deeper into the connections between the in-gap states and the sign-changing of the order parameter, we derive the $T$-matrix analytically. Since the spectral function and LDOS spectra are generally small at low energies, the imaginary part of $A(\omega)$ is also generally small within the superconducting gap. As a result, the impurity-induced low-energy features are mainly determined by the real part of $A(\omega)$. The real part of $A(\omega)$ at zero energy is expressed as:
\begin{eqnarray}
A(0) = V_{\text{imp}}^2 \left[ \left( \sum_{\mathbf{k}} P_{\mathbf{k}} \right)^2 + \left( \sum_{\mathbf{k}} Q_{\mathbf{k}} \right)^2 \right],
\end{eqnarray}
where $P_{\mathbf{k}}$ and $Q_{\mathbf{k}}$ can be written as:
\begin{eqnarray}
P_{\mathbf{k}} &=& \frac{\varepsilon_{\mathbf{k}}}{\varepsilon_{\mathbf{k}}^2 + \Delta_{\mathbf{k}}^2}, \\
Q_{\mathbf{k}} &=& \frac{\Delta_{\mathbf{k}}}{\varepsilon_{\mathbf{k}}^2 + \Delta_{\mathbf{k}}^2}.
\end{eqnarray}

$P_{\mathbf{k}}$ and $Q_{\mathbf{k}}$ are large only in the vicinity of the normal state Fermi surface and should rapidly drop to nearly zero away from the Fermi surface. The intensity plots of $P_{\mathbf{k}}$ and $Q_{\mathbf{k}}$ are presented in Figs.~\ref{fig1}(e) and \ref{fig1}(f), respectively. $P_\mathbf{k}$ is positive for the hole region and negative for the electron region, and it changes sign as it crosses the Fermi surface. For a half-filled system, the summation of $P_\mathbf{k}$ over the entire Brillouin zone results in positive and negative values canceling each other out, generally leading to a zero value.

For cuprate superconductors, there are two factors that lead to the summation of $P({\bf k})$ being close to zero value. First, superconductivity in cuprates is realized through doping Mott insulators, with the electron filling being close to the half-filling level, so the summation of $P({\bf k})$ is not too large. Second, the normal state band $\varepsilon_{\mathbf{k}}$ has a van Hove singularity at the momentum $(\pi,0)$. As a result, one generally expects that the Fermi surface near $(\pi,0)$ should mainly contribute to the summation of $P({\bf k})$. However, for cuprate superconductors with $d_{x^2-y^2}$ pairing, the gap magnitude reaches its maximum value at the $(\pi,0)$ position. Therefore, due to superconducting pairing, $P({\bf k})$ near $(\pi,0)$ is greatly suppressed, resulting in the summation of $P({\bf k})$ being close to zero over a wide range of doping concentrations.
 
On the other hand, for the $d$-wave pairing symmetry, both the normal state energy bands and the gap magnitudes have the $C_4$ rotational symmetry with $\varepsilon(k_x,k_y) \equiv \varepsilon(-k_y,k_x)$ and $\Delta(k_x,k_y) \equiv -\Delta(-k_y,k_x)$, thus the summation of $Q_\mathbf{k}$ is exactly zero. Consequently, $A(\omega)$ is nearly zero at the Fermi level for cuprate high-$T_c$ superconductors, leading to strong resonant peaks at the Fermi level in the LDOS spectrum near an impurity.

We now investigate whether an inevitable connection exists between the sign change of the order parameter along the normal state Fermi surface and in-gap states. We introduce an additional $s$-wave component into the $d_{x^2-y^2}$-wave pairing function, defined as $\Delta_{\mathbf{k}} = \Delta_s + \Delta_d(\cos k_x - \cos k_y)/2$, with $\Delta_d = 0.2$.  This component breaks the $C_4$ symmetry of the gap magnitudes. As $\Delta_s$ increases, the gap nodes shift from the diagonal direction towards the Brillouin zone boundary. Consequently, $\sum_{\mathbf{k}} Q(\mathbf{k})$ increases as $\Delta_s$ increases. A critical $s$-wave magnitude $\Delta_s^c$ can be defined, where the gap nodes shift exactly into the Brillouin zone boundary. When $\Delta_s$ equals or exceeds $\Delta_s^c$, the sign of the order parameter $\Delta_{\mathbf{k}}$ does not change along the entire normal state Fermi surface. $\Delta_s^c$ depends on the chemical potential and is expressed as $ \Delta_s^c  = \Delta_d \left( 1 - \left| \mu/4 \right| \right)$~\cite{supp}.

\begin{figure}
	\centering
	\includegraphics[width = 8.5cm]{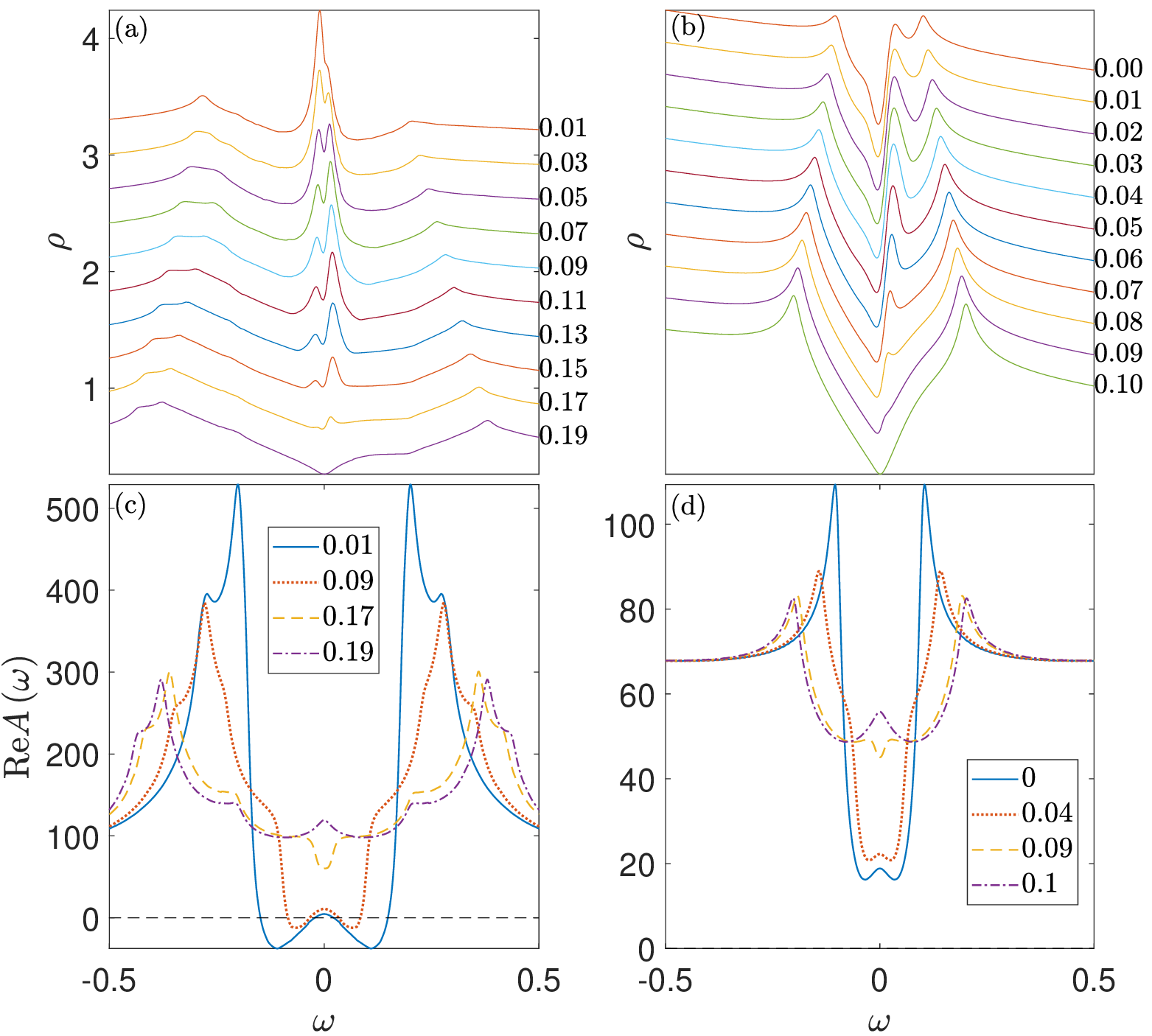}
	\caption{\label{fig2} (a) LDOS spectra at the nearest neighbor site of the impurity for the $s+d_{x^2 - y^2}$
wave model, with a chemical potential of $\mu=0.2$ and pairing defined as $\Delta_{\mathbf{k}} = \Delta_s + \Delta(\cos k_x - \cos k_y)/2$. The parameter $\Delta_s$
	increases from 0.01 to 0.19 from top to bottom. (b) Similar to panel (a), but for a chemical potential of $\mu=2$.
$\Delta_s$ increases from 0.01 to 0.1. Panels (c) and (d) show the real parts of the function 
 $A(\omega)$  corresponding to panels (a) and (b), respectively.
}
\end{figure}

We consider two different chemical potentials, $\mu=0.2$ and $\mu=2$, with corresponding critical $s$-wave magnitudes of $\Delta_s^c = 0.19$ and $\Delta_s^c = 0.1$, respectively. The LDOS spectra for different values of $\Delta_s$ are shown in Figs.~\ref{fig2}(a) and \ref{fig2}(b), and the corresponding real parts of the denominator of the $T$-matrix, $\text{Re } A(\omega)$, are presented in Figs.~\ref{fig2}(c) and \ref{fig2}(d). For the case of $\mu=0.2$, as seen from Fig.~\ref{fig2}(a), the presence of the $s$-wave component causes the mid-gap peak to split, with the intensity decreasing as $\Delta_s$ increases. The in-gap features disappear completely when $\Delta_s$ reaches the critical value (0.19). For the case of $\mu=2$, weak in-gap peaks appear at finite energy even in the absence of the $s$-wave component. Similarly, the in-gap features disappear completely when $\Delta_s$ reaches its corresponding critical value (0.1).

The in-gap features presented in Figs.~\ref{fig2}(a) and \ref{fig2}(b) are consistent with the denominator of the $T$-matrix, $\text{Re } A(\omega)$, shown in Figs.~\ref{fig2}(c) and \ref{fig2}(d). Specifically, when $\Delta_s$ is smaller than $\Delta_s^c$, $\text{Re } A(\omega)$ concaves down and has local minimal points at low energies, resulting in the in-gap feature. As $\Delta_s$ reaches the critical value, the curve of $\text{Re } A(\omega)$ concaves up, and the in-gap features disappear completely. 

Two key insights arise from these numerical results. First, the intensity of the resonant mid-gap peak decreases with increasing chemical potential. Second, there is a clear intrinsic relationship between the in-gap states and the sign reversal of the superconducting order parameter along the Fermi surface.

The absence of a strong resonant mid-gap peak for $\mu=2$ is attributed to heavy electron doping resulting from the large chemical potential. In this scenario, the negative part of $P_{\bf k}$ is significantly larger than the positive part. Additionally, the normal state Fermi surface shifts away from the van Hove singularity point,  leading a relatively larger negative value when summing $P_{\bf k}$ over the entire Brillouin zone. Consequently, the resonant condition $A(\omega)=0$ is not satisfied, resulting in weaker in-gap peaks where the real part of $A(\omega)=0$ is minimized, as shown in Fig. \ref{fig2}(d).

\begin{figure}
	\centering
	\includegraphics[width = 8.5cm]{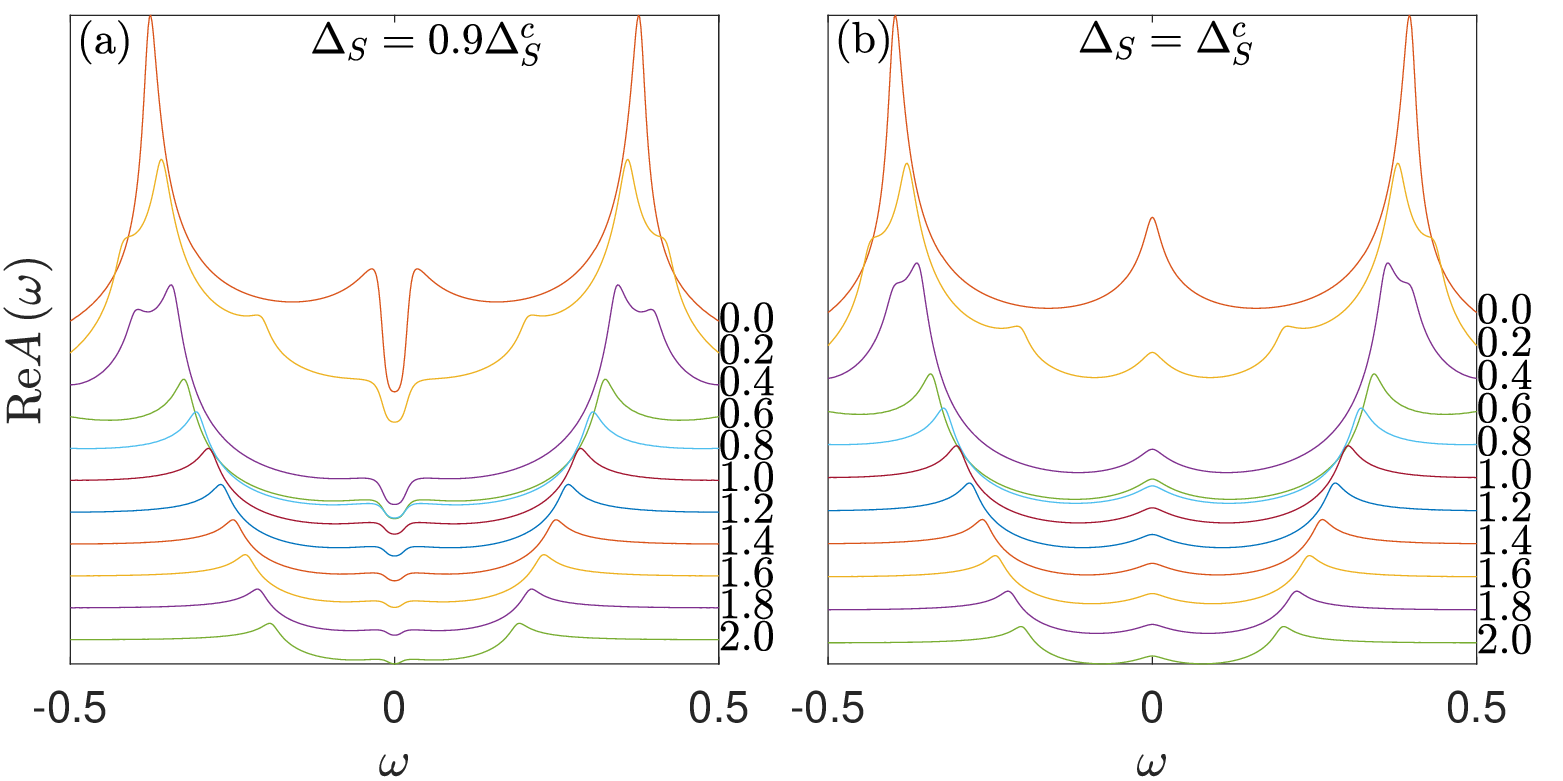}
	\caption{\label{fig3} (a) The real part of the function $A(\omega)$ as a function of $\omega$ for the $s+d_{x^2 - y^2}$ wave model with the $s$-wave component $\Delta_s = 0.9 \Delta^c_s$. The chemical potential $\mu$ increases from 0 to 2 from top to bottom. (b) Similar to panel (a) but for a different $s$-wave component $\Delta_s = \Delta^c_s$.
}
\end{figure}

The crucial connection between the in-gap states and the sign-changing of the order parameter is further confirmed and understood by exploring the denominator of the $T$-matrix as $\Delta_s$ crosses the critical value. We present the numerical results of $\text{Re } A(\omega)$ with $\Delta_s=0.9\Delta_s^c$ and $\Delta_s=\Delta_s^c$ for chemical potentials ranging from $0$ to $2$ in Figs.~\ref{fig3}(a) and \ref{fig3}(b), respectively. As is seen, when $\Delta_s$ is slightly below the critical value, the curve concaves down at low energies for all chemical potentials considered. Instead, as $\Delta_s$ increases to the critical value, all curves turn to concave up at low energies. When $\Delta_s$ is smaller than $\Delta_s^c$, the normal state Fermi surface has nodal points, and the order parameter changes sign crossing the nodal points. In this case, the summation of $Q_\mathbf{k}$ along the Fermi surface reduces because some positive and negative values of $Q_\mathbf{k}$ cancel each other out, leading to the concaving down behavior for $\text{Re } A(\omega=0)$. Such concaving down behavior further leads to the in-gap features.

We have established a fundamental connection between in-gap states and the sign change of the order parameter. Building on this insight, we can effectively harness impurity effects to probe the pairing symmetry of unconventional superconductors. Furthermore, we propose that impurity effects also serve as a powerful tool for investigating unconventional superfluid states in cold atom systems. In these systems, a point impurity can be simulated using a highly localized potential. Recently, experimental evidence for a chiral superfluid state was reported~\cite{Wang2021}. Thus, impurity effects can be further utilized to identify pairing states in this system, providing valuable insights into the nature of chiral superfluidity~\cite{supp}.

\section{Impurity effect in bilayer nickelate superconductor L$\mathrm{\bf a}_{\bf 2}$N$\mathrm{\bf i}_{\bf 3}$O$_{\bf 7}$ with the interlayer pairing}

The bilayer nickelate superconductor La$_3$Ni$_2$O$_7$ has recently attracted significant attention in the research community. Its low-energy electronic structure is primarily governed by the Ni-$d_{x^2-y^2}$ orbital and $d_{z^2}$ orbitals, with the latter proposed as a critical driver of superconductivity. The unique characteristics of the $d_{z^2}$
  orbital lead to a significantly enhanced interlayer hopping constant, which strengthens interlayer exchange interactions. This has motivated numerous studies proposing that interlayer pairing is the dominant mechanism driving superconductivity in this material \cite{rs.3.rs-3901266/v1, j.scib.2024.07.030, PhysRevB.108.L201108, PhysRevB.110.094509, PhysRevB.108.174511, PhysRevB.110.L041111, PhysRevLett.132.146002, 0256-307X/41/5/057403,PhysRevB.111.094504,huang2025spmpairing}.

Interlayer electron pairing is a unique property of Ruddlesden-Popper nickelate superconductors, and currently, there are no studies on the impurity effects in interlayer electron-paired superconductors. To address this gap, we begin with a simplified single-orbital model to investigate the role of impurities in interlayer electron-paired superconductors. 
 We focus on the bilayer system with strong interlayer hopping by setting \( l = 1,2 \) and \( t_{\perp} = 1.5 \) in Eq.~(1). The normal state Hamiltonian can be represented as a \(2 \times 2\) matrix. This matrix can be diagonalized by defining the following quasiparticle operators:
 \begin{eqnarray}
	\alpha_{{\bf k}\sigma}=&\frac{1}{\sqrt{2}}(c_{{\bf k}1\sigma}+c_{{\bf k}2\sigma}),\nonumber\\
	\beta_{{\bf k}\sigma}=&\frac{1}{\sqrt{2}}(c_{{\bf k}1\sigma}-c_{{\bf k}2\sigma}).
\end{eqnarray}
As a result, the normal state Fermi surface splits into two pockets defined by the equations \( \varepsilon_{\mathbf{k}} \pm t_{\perp} = 0 \).
 
 The superconducting order parameters in Eqs.~(2) and (3) are set as the $s$-wave inter-layer pairing with \( \Delta^{12}_{\mathbf{k}} = \Delta^{21}_{\mathbf{k}} = \Delta_{\perp} \). Similar to the case of intralayer pairing, we define the denominator of the $T$-matrix, $A(\omega)$, with the real part of $A(\omega)$ at zero energy being expressed as:
\begin{eqnarray}
A(0) &=& V_{\text{imp}}^2 \left[ \left( \sum_{\mathbf{k}} P'_{\mathbf{k}} \right)^2 + \left( \sum_{\mathbf{k}} Q'_{\mathbf{k}} \right)^2 \right], \\
P'_{\mathbf{k}} &=& \frac{\varepsilon_{\mathbf{k}} \left( \varepsilon_{\mathbf{k}}^2 - t_{\perp}^2 + \Delta_{\perp}^2 \right)}{D_H({\bf k})}, \\
Q'_{\mathbf{k}} &=& \frac{2 \varepsilon_{\mathbf{k}} t_{\perp} \Delta_{\perp}}{D_H({\bf k})}, \\
D_H({\bf k}) &=& t_{\perp}^4 + 2 t_{\perp}^2 \left( \Delta_{\perp}^2 - \varepsilon_{\mathbf{k}}^2 \right) + \left( \Delta_{\perp}^2 + \varepsilon_{\mathbf{k}}^2 \right)^2.
\end{eqnarray}

$P'_{\mathbf{k}}$ and $Q'_{\mathbf{k}}$ share some similar properties with $P_{\mathbf{k}}$ and $Q_{\mathbf{k}}$, i.e., $D_H({\bf k})$ reaches its minimum value as $\varepsilon_{\mathbf{k}}^2 = t_{\perp}^2$, thus both $P'_{\mathbf{k}}$ and $Q'_{\mathbf{k}}$ are large along the normal state Fermi surface and tend to zero away from the Fermi surface. $P'_{\mathbf{k}}$ changes sign as it crosses the Fermi surface, and the summation of $P'_\mathbf{k}$ over the entire Brillouin zone results in a very small value. $\Delta_{\perp}$ is a constant value, while $\varepsilon_{\mathbf{k}}$ changes sign for the two Fermi pockets. As a result, the summation of $Q'_{\mathbf{k}}$ over the entire Brillouin zone is generally small. Therefore, for superconductors with interlayer pairing, our analytical calculation indicates that in-gap resonant peaks generally exist.

\begin{figure}
	\centering
	\includegraphics[width = 8.5cm]{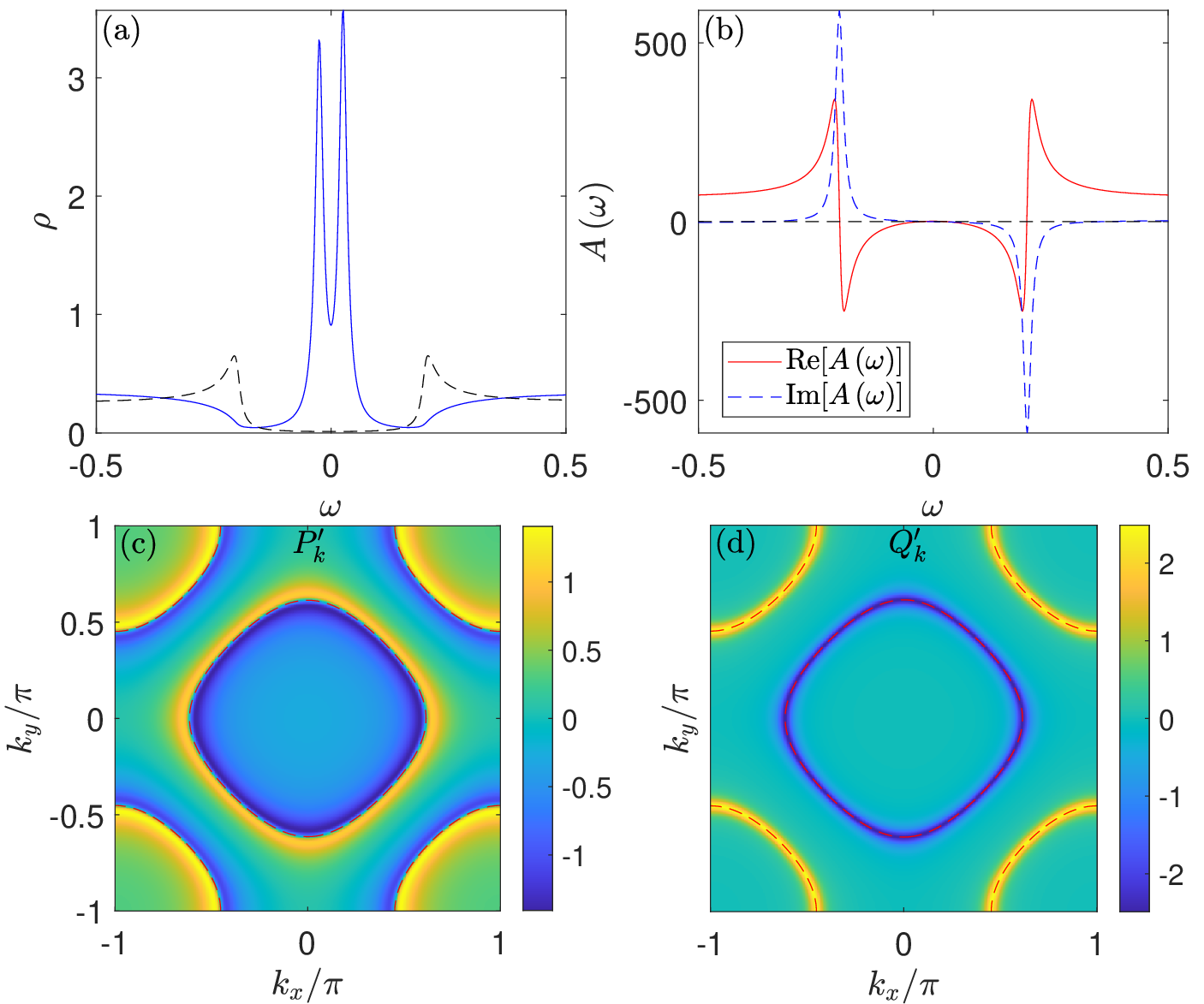}
	\caption{\label{fig4} (a) Solid line: LDOS spectrum at the nearest neighbor site of the impurity site for the interlayer pairing scenario. The dashed line represents the bare LDOS spectrum in the absence of the impurity. (b) Real and imaginary components of the function $A(\omega)$ for the interlayer pairing. Panels (c) and (d) display the intensity distribution maps for the functions $P'_{\mathbf{k}}$ and $Q'_{\mathbf{k}}$, respectively. The dashed lines in Panels (c) and (d) indicate the normal state Fermi surface.
	}
\end{figure}

This conclusion can be confirmed numerically. We present the numerical results of the LDOS spectra with $\Delta_{\perp} = 0.2$ and $\mu = 0.2$ in Fig.~\ref{fig4}(a). The real and imaginary parts of $A(\omega)$ as a function of $\omega$ are plotted in Fig.~\ref{fig4}(b). 
The intensity plots of functions $P'_{\mathbf{k}}$ and $Q'_{\mathbf{k}}$ are presented in Figs.~\ref{fig4}(c) and \ref{fig4}(d), respectively.
As is seen, without the impurity, the LDOS spectrum is 'U' shaped, indicating that the system is fully gapped. In the presence of an impurity, two sharp resonant peaks emerge, with the peak position lying symmetrically about the Fermi energy. The existence of the resonant peaks is consistent with the pole condition of the $T$-matrix. As seen in Fig.~\ref{fig4}(b), both the real part and the imaginary part of $A(\omega)$ tend to zero at low energies. This result is consistent with the analytical formulas of $A(\omega)$ presented in Eqs.~(12-15) and numerical results for the functions $P'_{\mathbf{k}}$ and $Q'_{\mathbf{k}}$ presented in Figs.~\ref{fig4}(c) and \ref{fig4}(d).

The existence of in-gap resonant states in superconductors with interlayer pairing can be coherently explained using the sign-reversal scenario presented in Sec. III. By substituting Eq. (11) into Eq. (2), the interlayer superconducting pairing term can be rewritten as:
\begin{equation}
	H_{\Delta} = \sum_{\bf k} \left( \Delta_\perp \alpha^\dagger_{{\bf k}\uparrow}\alpha^\dagger_{-{\bf k}\downarrow} - \Delta_\perp \beta^\dagger_{{\bf k}\uparrow}\beta^\dagger_{-{\bf k}\downarrow} + \text{H.c.} \right).
\end{equation}
In the band representation, the pairing function exhibits an $s_{\pm}$ pairing symmetry. The pairing order parameters for the two quasiparticle bands are exactly opposite in sign. Consequently, the presence of in-gap resonant states is generally expected.

\begin{figure}
	\centering
	\includegraphics[width = 8.5cm]{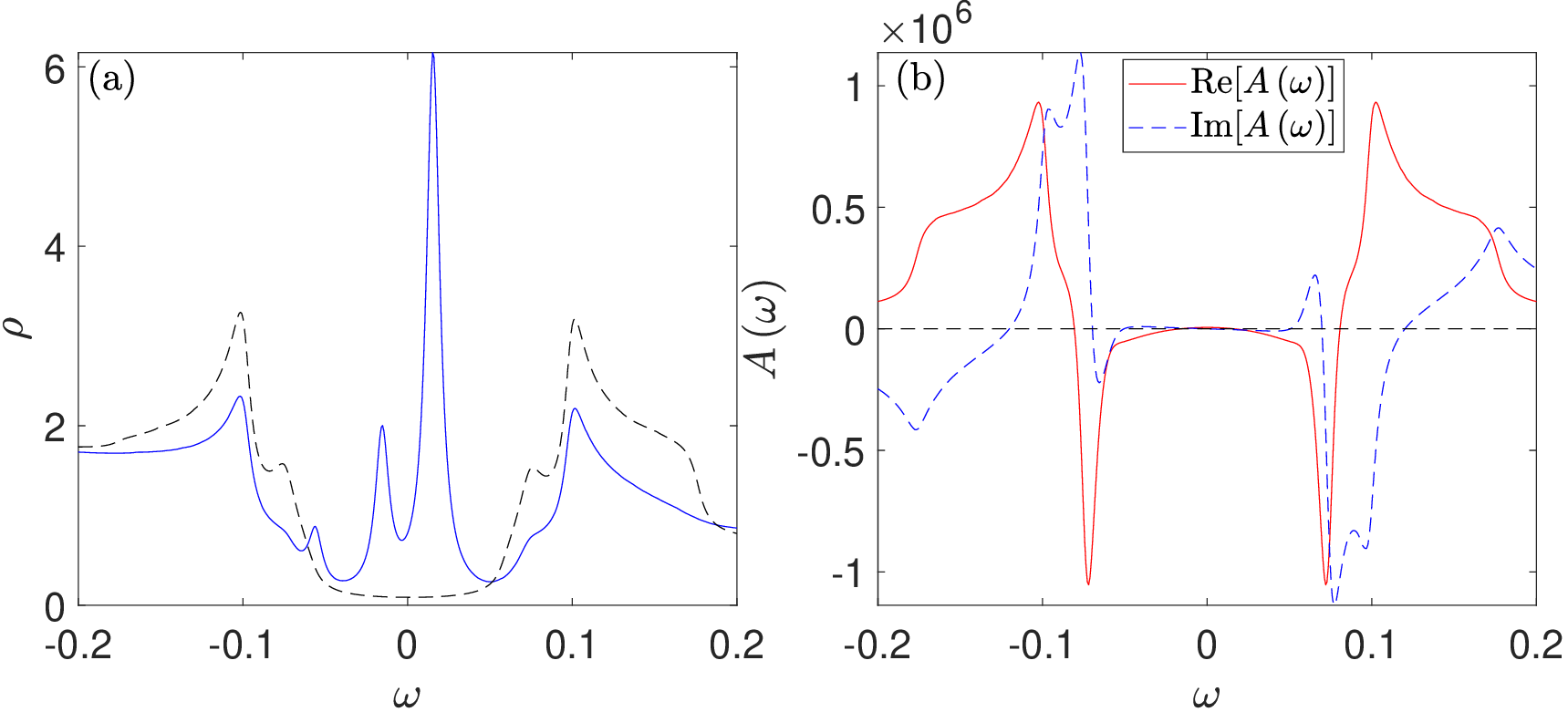}
	\caption{\label{fig5} (a) Solid line: LDOS spectrum at the nearest neighbor site of the impurity site for the nickelate superconductors with interlayer pairing scenario. The dashed line represents the bare LDOS spectrum in the absence of the impurity. (b) Real and imaginary components of the function $A(\omega)$ for the interlayer pairing. }
\end{figure}

We now adopt a more realistic model for La$_3$Ni$_2$O$_7$ to investigate its impurity effects. Based on density functional theory calculations, this material can be qualitatively described by a two-orbital bilayer model. The Hamiltonian in the superconducting state is represented as an \(8 \times 8\) matrix~\cite{supp}. 
 Recent numerical studies have confirmed that interlayer pairing within this model gives rise to a typical $s_{\pm}$ pairing symmetry~\cite{PhysRevB.111.094504,huang2025spmpairing}.
Fig.~\ref{fig5}(a) presents numerical results for the LDOS spectra, while Fig.~\ref{fig5}(b) illustrates the real and imaginary components of the function \(A(\omega)\) as a function of energy \(\omega\). Notably, the introduction of an impurity induces two distinct resonant peaks symmetrically positioned about the Fermi energy. This phenomenon is consistent with the pole condition of the \(T\)-matrix formalism. As depicted in Fig.~\ref{fig5}(b), both the real and imaginary components of \(A(\omega)\) exhibit a suppression at low energies, approaching zero as \(\omega \to 0\). These findings align qualitatively with results from the simplified bilayer model incorporating interlayer pairing presented in Fig.~\ref{fig4}.

We have presented the numerical results for single impurity effect for La$_3$Ni$_2$O$_7$ material with the interlayer pairing. 
Now the pairing symmetry for La$_3$Ni$_2$O$_7$ material remains an open question. Although the interlayer pairing has been proposed theoretically \cite{rs.3.rs-3901266/v1, j.scib.2024.07.030, PhysRevB.108.L201108, PhysRevB.110.094509, PhysRevB.108.174511, PhysRevB.110.L041111,PhysRevLett.132.146002, 0256-307X/41/5/057403,PhysRevB.111.094504,huang2025spmpairing}, the possibility of dominant intralayer still cannot be excluded.  
For intralayer dominant pairing, several possible pairing symmetries have been proposed theoretically, including the $s_{\pm}$ pairing symmetry and $d$-wave pairing symmetry~\cite{Wang_2024,2024arXiv241211429G}. It has also been proposed that the dominant interlayer pairing accounts for the superconductivity Our impurity scattering theory may be directly applied to nickelate superconductors and used to probe the pairing function. Based on the numerical results presented in Sec. III, we infer that if intra-layer $d$-wave pairing is predominant, strong mid-gap states should be present. Conversely, if intra-layer $s_{\pm}$-wave pairing is dominant, the positive and negative contributions to $Q_{\mathbf{k}}$ cannot be completely cancelled out, leading to the presence of weaker in-gap states. The numerical results for these two pairing symmetries, which are in agreement with our conclusions, have been detailed in Ref. \cite{PhysRevB.108.174501}.

Experimental investigations of impurity scattering in the superconducting state of La$_3$Ni$_2$O$_7$ have yet to be conducted. 
 The initial discovery of this material required high-pressure conditions to induce superconductivity~\cite{s41586-023-06408-7}, posing challenges for further measurements of its physical properties. However, recent breakthroughs have enabled superconductivity in La$_3$Ni$_2$O$_7$ and La$_{2.85}$Pr$_{0.15}$Ni$_2$O$_7$ thin films at ambient pressure~\cite{s41586-024-08525-3,Zhou2025}. This development opens new avenues for exploring the physical properties of these materials under more accessible experimental conditions.
Angle-resolved photoemission spectroscopy measurements of the superconducting gap in La$_{2.85}$Pr$_{0.15}$Ni$_2$O$_7$ thin films have been reported~\cite{arXiv2502.17831}, and the results could hint at an interlayer pairing mechanism~\cite{PhysRevB.111.094504,huang2025spmpairing}. Future STM are expected to provide experimental observation for impurity effect and allow for direct comparison with theoretical calculations, further advancing our understanding of this intriguing system.

This work provides valuable insights into the effects of a point impurity in unconventional superconductors and establishes a clear link between bound states and the sign change of the superconducting order parameter. However, several critical challenges remain to be addressed in future research.
First, the original Hamiltonian considered here is a non-interacting model. Incorporating electronic correlations and examining how interaction terms influence the formation and properties of bound states could deepen our understanding of impurity effects in these systems.
Second, the impurity is treated as a point potential in this study. Extending the impurity model to alternative forms, such as Anderson impurities, may yield important new insights into how impurity characteristics affect superconductivity.
Moreover, investigating systems with finite impurity concentrations and the associated disorder effects is crucial. In such scenarios, phenomena like Anderson localization could emerge, significantly impacting the electronic and superconducting properties of the material. Addressing these challenges will be essential for advancing our comprehension of impurity effects in unconventional superconductors.

\section{\label{sec:Summary}Summary}
We have undertaken a detailed exploration of single impurity scattering in unconventional superconductors, focusing on intralayer $d$-wave and interlayer $s$-wave pairing. Specifically, in the case of intralayer $d_{x^2-y^2}$-wave pairing near half-filling doping in a square lattice, our study reveals an inherent connection between the $d_{x^2-y^2}$-wave pairing symmetry and the emergence of mid-gap states. Furthermore, we establish a significant interplay between in-gap states and the sign change of the order parameter.
In interlayer pairing systems, prominent resonant peaks are revealed, which can also be attributed to the sign-changing nature of the pairing order parameter. By employing the $T$-matrix approach, we successfully elucidate the underlying mechanisms responsible for these impurity-induced states.
This theoretical framework proves highly applicable to the analysis of newly discovered bilayer nickel-based high-temperature superconductor La$_3$Ni$_2$O$_7$, providing a valuable means for distinguishing its pairing properties.

Overall, this study illuminates the intricate relationship between pairing symmetries and impurity effects in unconventional superconductors.  It establishes a robust foundation for future research on the unique characteristics of these materials and offers an effective method for probing the pairing symmetry of the bilayer nickel-based superconductor La$_3$Ni$_2$O$_7$.


	This work was supported by the NSFC under the Grant No.12074130.
%

\renewcommand{\thesection}{S-\arabic{section}}
\setcounter{section}{0}  
\renewcommand{\theequation}{S\arabic{equation}}
\setcounter{equation}{0}  
\renewcommand{\thefigure}{S\arabic{figure}}
\setcounter{figure}{0}  
\renewcommand{\thetable}{S\Roman{table}}
\setcounter{table}{0}  
\onecolumngrid \flushbottom 
\newpage
\begin{center}\large \textbf{Supplemental Material For Probing Sign-Changing Order Parameters via Impurity States in unconventional superconductors: Implications for La$_3$Ni$_2$O$_7$ Superconductors with interlayer pairing} \end{center}

\section{the critical $s$-wave component}

\renewcommand \thefigure {S\arabic{figure}}
\begin{figure*}[htb]
\centering
\includegraphics[width = 10cm]{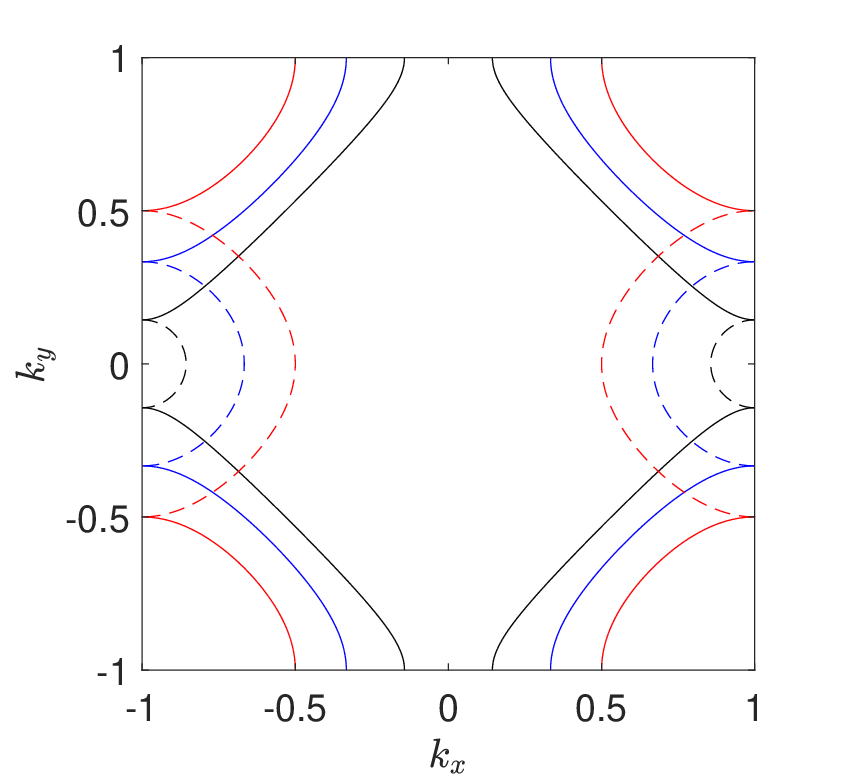}
\caption{\label{figS1} The solid red, blue, and black lines represent the Fermi surfaces in the normal state for chemical potentials of \(\mu = 2\), \(\mu = 1\), and \(\mu = 0.2\), respectively. The corresponding dashed red, blue, and black lines depict the nodal lines of the energy gap when \(\Delta_s = \Delta^c_s(\mu)\).}
\end{figure*}

In the main text, for the $s+d_{x^2-y^2}$ pairing symmetry, the order parameter is given by $\Delta_{\mathbf{k}} = \Delta_s + \Delta_d(\cos k_x - \cos k_y)/2$. In this context, we define the critical $s$-wave component $\Delta^c_s$, which depends on the chemical potential $\mu$ according to the equation $\Delta_s^c(\mu) = \Delta_d \left( 1 - \left| \mu/4 \right| \right)$. At the critical value $\Delta_s = \Delta^c_s$, the normal state Fermi surface is tangent to the nodal line of the energy gap, resulting in no sign change of the order parameter along the Fermi surface. To more intuitively illustrate this point, Fig. S1 presents the Fermi surfaces in the normal state for different chemical potentials $\mu$, along with the corresponding nodal lines of the energy gap when $\Delta_s = \Delta^c_s(\mu)$. As observed in the figure, the nodal lines of the energy gap intersect with the Fermi surface at the boundaries of the Brillouin zone, ensuring that there is no sign change of the energy gap across the entire normal state Fermi surface.

\section{Impurity effect of $p+ip$-wave pairing system}

\renewcommand \thefigure {S\arabic{figure}}
\begin{figure*}[htb]
\centering
\includegraphics[width = 12cm]{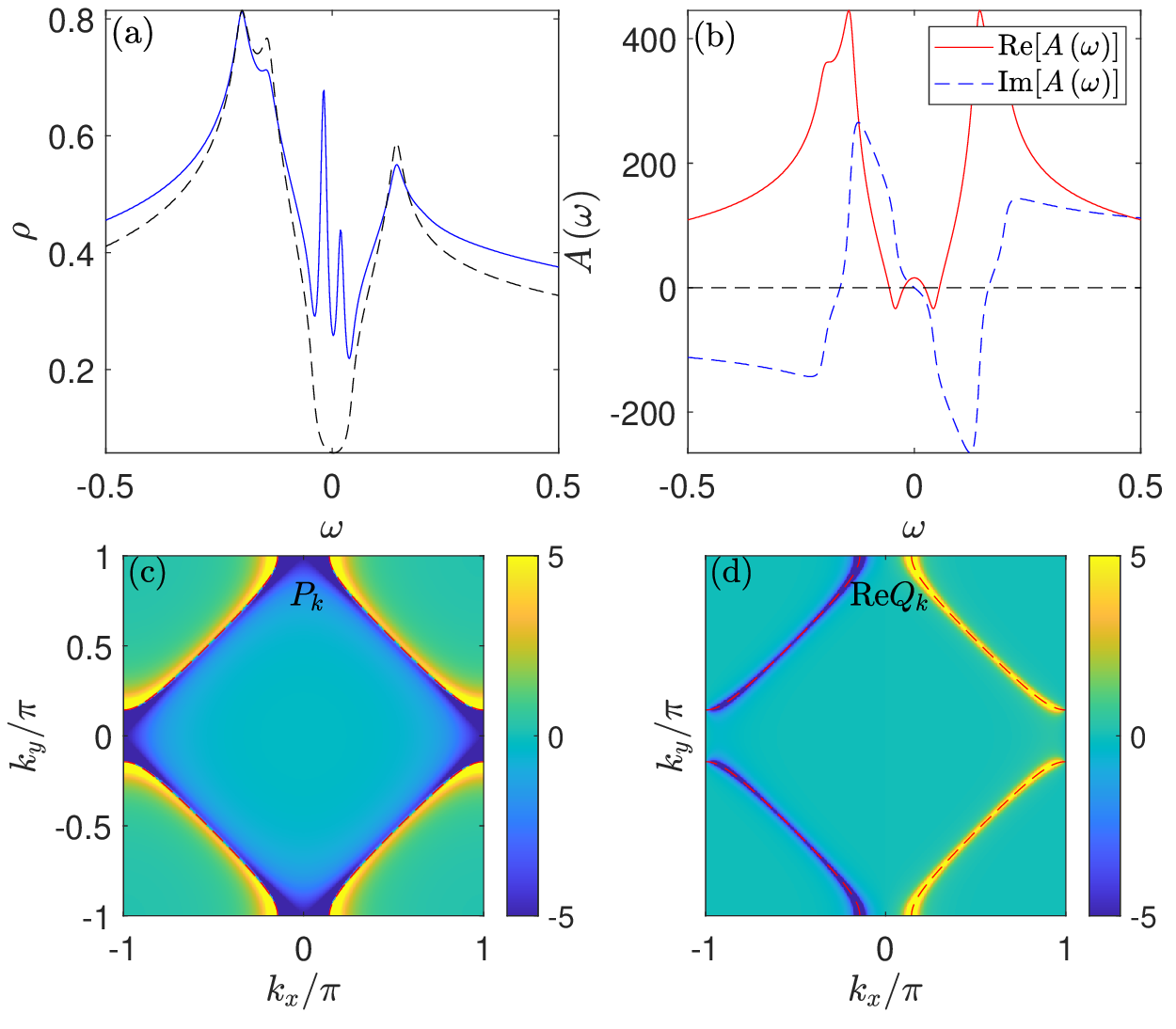}
\caption{\label{figS2} (a) Solid line: LDOS spectrum at the nearest neighbor site of the impurity site for a $p+ip$-wave pairing scenario. The dashed line represents the bare LDOS spectrum in the absence of the impurity. (b) Real and imaginary components of the function $A\left(\omega\right)$ for the $d$-wave pairing symmetry. Panels (c) and (d) display the intensity distribution maps for the functions $P_{\bf k}$ and Re$Q_k$, respectively. The dashed lines in Panels (c) and (d) indicate the normal state Fermi surface.}
\end{figure*}

In the main text, we have mentioned that the impurity effect can also be used to probe the chiral pairing states. We now present the numerical results of the $p+ip$ pairing state. 
In the $p+ip$-wave pairing scenario, the order parameter is given by $\Delta_{\bf k} = \Delta_0\left(\sin {\bf k_x} + i\sin {\bf k_y}\right)$. The real part of $A\left(\omega\right)$ at zero energy is expressed as:
\begin{eqnarray}
A(0) = V_{\text{imp}}^2 \left[ \left( \sum_{\mathbf{k}} P_{\mathbf{k}} \right)^2 + \left( \sum_{\mathbf{k}} Q_{\mathbf{k}} \right)^2 \right],
\end{eqnarray}
where $P_{\mathbf{k}}$ and $Q_{\mathbf{k}}$ are defined as:
\begin{eqnarray}
P_{\mathbf{k}} &=& \frac{\varepsilon_{\mathbf{k}}}{\varepsilon_{\mathbf{k}}^2 + \left|\Delta\right|_{\mathbf{k}}^2}, \\
Q_{\mathbf{k}} &=& \frac{\Delta_{\mathbf{k}}}{\varepsilon_{\mathbf{k}}^2 + \left|\Delta\right|_{\mathbf{k}}^2}.
\end{eqnarray}

We present the numerical results for the impurity effect in a $p+ip$ pairing system. The LDOS spectrum at the nearest-neighbor site of an impurity is shown in Fig.~\ref{figS2}(a). The corresponding numerical calculations for the real and imaginary parts of the function $A(\omega)$ are displayed in Fig.~\ref{figS2}(b).
The intensity plots of $P_{\mathbf{k}}$ and the real part of $Q_{\mathbf{k}}$ are presented in FIGs. \ref{figS2}(c) and \ref{figS2}(d), respectively. As is seen, two sharp impurity-induced resonance peaks exist in the LDOS spectrum.  
This phenomenon can be further understood by analyzing the functions  $P_{\mathbf{k}}$ and $Q_{\mathbf{k}}$. As is seen from FIGs. \ref{figS2}(c) and \ref{figS2}(d),
$P_{\mathbf{k}}$ at the two sides of Fermi surface changes sign and cancels each other, resulting in a small value. For the $p$-wave paring symmetry, both the normal state energy bands and the gap magnitudes have the $C_2$ rotational symmetry with $\varepsilon \left( k_x, k_y \right) \equiv \varepsilon \left( -k_x, -k_y \right)$ and $\Delta \left( k_x, k_y \right) \equiv - \Delta \left( -k_x, -k_y \right)$. Therefore, the sum of $Q_{\mathbf{k}}$ is exactly zero. As a result,$A(\omega)$ exhibits a pronounced suppression at zero energy, as shown in Fig.~\ref{figS2}(b). This suppression leads to the formation of two sharp impurity-induced resonance peaks in the LDOS spectrum near the impurity, as depicted in Fig.~\ref{figS2}(a). The numerical results further corroborate the relationship between impurity-induced states and the sign-changing nature of the order parameter.

\section{Model and Formalism for the L${\bf \mathrm{\bf a}_2}$N$\mathrm{\bf i}_{\bf 3}$O$_{\bf 7}$ material}

In the main text, we have presented the numerical results the single impurity effect in nickelate superconductors with interlayer $s$-wave pairing. The Hamiltonian can be written as $\hat{H}_{Ni} = \sum_{\bf k} \hat{\Psi}_{\bf k}^{\dagger} \hat{H}_{Ni}\left({\bf k}\right) \hat{\Psi}_{\bf k}$. $\hat{H}_{Ni}\left({\bf k}\right)$ is an $8 \times 8$ matrix expressed as
\begin{eqnarray}
\hat{H}_{Ni}\left({\bf k}\right) = \left( \begin{array}{cc}
{\hat{H}_t\left({\bf k}\right)}&{\hat{H}_{\Delta}\left({\bf k}\right)}\\
{\hat{H}_{\Delta}^{\dagger}\left({\bf k}\right)}&{-\hat{H}_t\left({\bf k}\right)}
\end{array} \right).
\end{eqnarray}
$\hat{H}_t\left({\bf k}\right)$ is the $4\times 4$ matrix, representing the normal state part,
\begin{eqnarray}
\hat{H}_{t}\left({\bf k}\right) = \left( \begin{array}{cc}
{\hat{H}_A\left({\bf k}\right)}&{\hat{H}_{AB}\left({\bf k}\right)}\\
{\hat{H}_{AB}\left({\bf k}\right)}&{\hat{H}_A\left({\bf k}\right)}
\end{array} \right),
\end{eqnarray}
where
\begin{eqnarray}
\hat{H}_A\left({\bf k}\right) = \left( \begin{array}{cc}
{{T}_{x{\bf k}}}&{{V}_{\bf k}}\\
{V_{\bf k}}&{T_{z{\bf k}}}
\end{array} \right) 
,
H_{AB}\left({\bf k}\right) = \left( \begin{array}{cc}
{t_{x\perp}}&{V'_k}\\
{V'_k}&{t_{z\perp}}
\end{array} \right).
\end{eqnarray}
Here
\begin{eqnarray}
T_{x{\bf k}} &=& 2t_{1x}\left( \cos { k_x} + \cos { k_y} \right) + 4t_{2x} \cos { k_x} \cos { k_y} + \epsilon_x \\
T_{z{\bf k}} &=& 2t_{1z}\left( \cos { k_x} + \cos { k_y} \right) + 4t_{2z} \cos { k_x} \cos { k_y} + \epsilon_z \\
V_{\bf k} &=& 2t_{3xz}\left( \cos { k_x} - \cos { k_y} \right) \\
V'_{\bf k} &=& 2t_{4xz}\left( \cos { k_x} - \cos { k_y} \right)
\end{eqnarray}
The tight-binding parameters referenced in \cite{PhysRevLett.131.126001} are as follows:

\begin{table}[h]
\centering
\begin{tabular}{|c|c|c|c|c|c|c|c|c|c|}
\hline
\multicolumn{1}{|c|}{$t_{1x}$} & \multicolumn{1}{c|}{$t_{1z}$} & \multicolumn{1}{c|}{$t_{2x}$} & \multicolumn{1}{c|}{$t_{1z}$} & \multicolumn{1}{c|}{$t_{3xz}$} & \multicolumn{1}{c|}{$t_{x\perp}$} & \multicolumn{1}{c|}{$t_{z\perp}$} & \multicolumn{1}{c|}{$t_{4xz}$} & \multicolumn{1}{c|}{$\epsilon_x$} & \multicolumn{1}{c|}{$\epsilon_z$} \\ \hline
-0.483 & -0.110 & 0.069 & -0.017 & 0.239 & 0.005 & -0.635 & -0.034 & 0.776 & 0.409 \\ \hline
\end{tabular}
\end{table}

 $\hat{H}_{\Delta}$ is the superconducting pairing order part of Hamiltonian. Consiering $s$-wave interlayer pairing, $\hat{H}_{\Delta}$ is expressed as
\begin{eqnarray}
\hat{H}_{\Delta}\left({\bf k}\right) = \left( \begin{array}{cccc}
{0}&{0}&{\Delta_{x\perp}}&{0}\\
{0}&{0}&{0}&{\Delta_{z\perp}}\\
{\Delta_{x\perp}}&{0}&{0}&{0}\\
{0}&{\Delta_{z\perp}}&{0}&{0}\\
\end{array} \right),
\end{eqnarray}
The base vector is $\hat{\Psi}_{\bf k}^{\dagger} = \left( c_{{\bf k}1x\uparrow}^{\dagger}, c_{{\bf k}1z\uparrow}^{\dagger}, c_{{\bf k}2x\uparrow}^{\dagger}, c_{{\bf k}2z\uparrow}^{\dagger}, c_{{\bf k}1x\downarrow}, c_{{\bf k}1z\downarrow}, c_{{\bf k}2x\downarrow}, c_{{\bf k}2z\downarrow} \right)$. Here the subscripts $1,2$ represent the layer and the subscripts $x,z$ represent the orbital. 

The interlayer superconducting order parameters are determined self-consistently as
\begin{eqnarray}
\Delta_{x/z \perp} = \frac{V}{2N} \sum_{n{\bf k}} u_{x/z,n{\bf k}}^{*}v_{x/z,n{\bf k}} \tanh \frac{\beta E_{n{\bf k}}}{2},
\end{eqnarray}
where $V$ is the pairing potential with $V=0.8$ being considered. 

The bare Green's function matrix for a clean system can be obtained through diagonalizing the Hamiltonian matrix, with the elements being defined as
\begin{eqnarray}
	G_{0ij}\left({\bf k},\omega\right) =\sum_n \frac{u_{in}({\bf k})u^*_{jn}({\bf k})}{\omega - E_n\left({\bf k}\right) + i\Gamma},
\end{eqnarray}
where $u_{in}({\bf k})$ and $E_n({\bf k})$ are the eigenvectors and the eigenvalue
of the Hamiltonian matrix, respectively. $\Gamma$ is a small constant.

In the main text, we consider a single impurity at the site $\left(0,\space 0\right)$ of layers 1, the $T$-matrix can be expressed as
\begin{eqnarray}
	\hat{T}\left(\omega\right) = \frac{\hat{U}}{\hat{I} - \hat{U}\hat{G}_{0}\left({\bf r},{\bf r},\omega\right)},
\end{eqnarray}
where \(\hat{I}\) is the identity matrix, \(\hat{U}\) is a diagonal matrix with non-zero elements \(U_{11} = U_{22} = V_{\text{imp}}\) and \(U_{55} = U_{66} = -V_{\text{imp}}\), and \(\hat{G}_{0}\left(\mathbf{r},\mathbf{r'},\omega\right)\) is the Fourier transform of \(\hat{G}_0\left(\mathbf{k},\omega\right)\) with
$
\hat{G}_{0}\left(\mathbf{r},\mathbf{r'},\omega\right) = \frac{1}{N} \sum_{\mathbf{k}} \hat{G}_0\left(\mathbf{k},\omega\right) e^{i\mathbf{k} \cdot (\mathbf{r}-\mathbf{r'})}$.

The full Green's function in the presence of a single impurity is calculated by:
\begin{equation}
\hat{G}\left(\mathbf{r},\mathbf{r'},\omega\right) = \hat{G}_0\left(\mathbf{r},\mathbf{r'},\omega\right) + \hat{G}_0\left(\mathbf{r},0,\omega\right) \hat{T}\left(\omega\right) \hat{G}_0\left(0,\mathbf{r'},\omega\right).
\end{equation}

The LDOS at layer \(l\) and site \(\mathbf{r}\) can be calculated using the full Green's function:
\begin{equation}
\rho_{l}\left(\mathbf{r},\omega\right) = -\frac{1}{\pi} \text{Im} \sum_{p=0}^1 \left[ G_{l+p,l+p}\left(\mathbf{r},\mathbf{r},\omega\right) + G_{l+p+4,l+p+4}\left(\mathbf{r},\mathbf{r},-\omega\right) \right].
\end{equation}


\end{document}